\documentclass[aps,pra,preprint,superscriptaddress,longbibliography,floatfix]{revtex4-2}

\usepackage[T1]{fontenc}
\usepackage{amsmath,amssymb,bm}
\usepackage{booktabs}
\usepackage{graphicx}
\usepackage{hyperref}
\usepackage{xcolor}


\newcommand{\uu}{\bm{u}}
\newcommand{\zz}{\bm{z}}
\newcommand{\yy}{\bm{y}}
\newcommand{\Jfid}{\mathcal{J}_{\mathrm{fid}}}

\begin{document}
\raggedbottom

\title{Constraint-native quantum control for fidelity--complexity trade-offs with inexact proximal ADMM}

\author{Ziwen Song}
\affiliation{Jilin University, College of Instrumentation and Electrical Engineering, Changchun, People's Republic of China}
\email{songzw24@mails.jlu.edu.cn}

\begin{abstract}
Quantum-control pulses are often optimised for nominal fidelity before waveform constraints are imposed. This sequence can conceal the fidelity cost of producing smooth, band-limited, and amplitude-admissible controls. Here, we evaluate a constraint-native alternative based on inexact proximal alternating-direction updates. The formulation combines gate-infidelity minimisation with amplitude bounds, Fourier-domain bandwidth projection, amplitude sparsity, and total-variation regularisation. We compare it with GRAPE, standard Krotov optimisation, and L-BFGS-B on a single-qubit gate, a leakage-prone qutrit gate, and a two-qubit entangler without a directly controlled target generator. Random seeds are paired across methods, and qutrit computational-subspace fidelity is reported alongside leakage. PADMM-Warm reached mean qutrit and two-qubit fidelities of $0.6363$ and $0.9541$, respectively, while reducing total variation by factors of $13.2$ and $10.7$ relative to L-BFGS-B. These results define a reproducible fidelity--complexity trade-off, not a universal fidelity advantage. The method is therefore a numerical tool for exploring low-complexity control frontiers rather than a replacement for unconstrained high-fidelity solvers.
\end{abstract}

\maketitle

\section{Introduction}

Quantum optimal control provides numerical procedures for steering driven systems towards target states or gates
\cite{Rabitz2000,Werschnik2007,Brif2010,DAlessandro2008}.
Gradient-based methods, including GRAPE and Krotov optimisation, are widely used across quantum information, atomic physics, and molecular control
\cite{Palao2003,Khaneja2005,Goerz2019,Glaser2015,Machnes2011,VanFrank2016}.
Quasi-Newton methods can also achieve high nominal fidelities when the pulse parameterisation remains manageable
\cite{deFouquieres2011,Nocedal2006}.

Nominal fidelity is not the only property that determines whether a pulse is useful.
Large amplitudes, rapid temporal variation, and spectral weight outside an available control band can hinder interpretation and implementation
\cite{Kosloff1989,Shi1988,Somloi1993,Rach2015,Koch2016,Goerz2014,PalaoReichKoch2013}.
These restrictions are particularly relevant in multilevel and entangling systems, where leakage suppression and nontrivial gate synthesis require structured control resources
\cite{Motzoi2009,Schutjens2013,Nebendahl2009,Muller2011}.

A common workflow first optimises fidelity and then filters or smooths the waveform.
Such post-processing can reduce complexity, but it evaluates a waveform different from the one selected by the original objective.
The resulting fidelity penalty is therefore absent from the optimisation decision.
Alternative parameterisations and landscape-aware searches address related resource constraints, but do not remove the need to report the associated trade-off explicitly
\cite{Caneva2011,Doria2011,Chakrabarti2007}.
Constraint-native optimisation instead places admissibility inside the optimisation loop.

This work examines an inexact proximal alternating-direction method of multipliers (PADMM) for that purpose.
The method uses variable splitting to separate amplitude sparsity, temporal total variation, and Fourier bandwidth
\cite{Boyd2011,Parikh2014}.
It is complementary to hybrid and calibration-aware control strategies, which combine optimisation stages or use experimental feedback
\cite{Eitan2011,Goerz2015,Machnes2018,Egger2014,Kelly2014}.
The control-variable subproblem is updated by a finite number of gradient steps, so the method is not textbook ADMM.
Our question is deliberately narrow: can this split formulation identify reproducible low-complexity pulses across several gate-synthesis tasks under a fair, seed-paired comparison?

The revised study incorporates four methodological safeguards that make the comparison auditable.
First, the two-qubit task no longer contains a direct control proportional to the target generator.
Second, qutrit performance is evaluated on the computational subspace and accompanied by an explicit leakage measure.
Third, robust training and testing use the same amplitude-limited perturbation map.
Fourth, statistical comparisons are paired by random seed and robustness levels are aggregated within each seed.

\section{Model and methods}

\subsection{Piecewise-constant dynamics}

We consider closed-system dynamics with drift Hamiltonian $H_0$ and control Hamiltonians $H_m$,
\begin{equation}
 \dot U(t)=-i\left[H_0+\sum_{m=1}^{M}u_m(t)H_m\right]U(t),
 \qquad U(0)=I.
\end{equation}
We divide the total duration into $N$ intervals of width $\Delta t$.
For piecewise-constant controls, one propagation step is
\begin{equation}
 U_{k+1}=
 \exp\left\{-i\Delta t\left[H_0+\sum_m u_m[k]H_m\right]\right\}U_k.
\end{equation}
The final propagator is the ordered product of these step propagators.

For a $d$-dimensional target $U_{\mathrm{tar}}$, the full-space trace fidelity is
\begin{equation}
 F_{\mathrm{full}} =
 \left|\frac{1}{d}\operatorname{Tr}
 \left(U_{\mathrm{tar}}^\dagger U_N\right)\right|^2.
\end{equation}
The fidelity objective is $\Jfid=1-F_{\mathrm{full}}$.
Its exact slice derivatives are evaluated with Fr\'echet derivatives of the matrix exponential.
Finite-difference checks are applied to every benchmark model.

\subsection{Qutrit subspace fidelity and leakage}

Let $P=[|0\rangle,|1\rangle]$ embed the computational subspace in the qutrit space.
The projected target and evolution are $U_{\mathrm{tar},c}=P^\dagger U_{\mathrm{tar}}P$ and $U_c=P^\dagger U_NP$.
We report
\begin{equation}
 F_c=\left|\frac{1}{2}\operatorname{Tr}
 \left(U_{\mathrm{tar},c}^\dagger U_c\right)\right|^2
\end{equation}
as the primary qutrit metric.
This quantity penalises logical error and population loss.
We separate the latter through the mean leakage
\begin{equation}
 L=1-\frac{1}{2}\operatorname{Tr}(U_c^\dagger U_c).
\end{equation}
The full-space fidelity remains a secondary diagnostic.

\subsection{Benchmark Hamiltonians}

The single-qubit task uses
\begin{equation}
 H_0=\frac{\omega}{2}\sigma_z,\qquad
 H_x=\frac{1}{2}\sigma_x,\qquad H_y=\frac{1}{2}\sigma_y,
\end{equation}
with target $\sigma_x$.

The qutrit model uses a weakly anharmonic ladder,
\begin{equation}
 H_0=\operatorname{diag}(0,\omega_{01},2\omega_{01}+\alpha),
\end{equation}
and quadrature drives with adjacent-level matrix elements in the ratio $1:\sqrt{2}$.
The target exchanges $|0\rangle$ and $|1\rangle$ while leaving $|2\rangle$ unchanged.

The two-qubit model contains local $x$ and $y$ controls on both qubits and a fixed $ZZ$ coupling,
\begin{align}
 H_0={}&\frac{\omega_1}{2}Z\otimes I+\frac{\omega_2}{2}I\otimes Z
 +\frac{J}{4}Z\otimes Z.
\end{align}
The target is $\exp[-i(\pi/4)X\otimes X]$.
No $X\otimes X$ control channel is included.

\subsection{Structured objective and split updates}

The structured problem is
\begin{equation}
\begin{split}
 \min_{\uu}\quad&
 \Jfid(\uu)+\lambda_1\|\uu\|_1+\lambda_{\mathrm{TV}}\|D\uu\|_1\\
 &+\delta_{\mathcal A}(\uu)+\delta_{\mathcal B}(\uu),
\end{split}
\label{eq:structured}
\end{equation}
where $D$ is the first-difference operator.
The set $\mathcal A$ imposes elementwise amplitude bounds.
The set $\mathcal B$ contains controls whose discrete Fourier coefficients vanish above a specified cutoff.

We introduce splits $\zz_1=\uu$, $\zz_2=D\uu$, and $\zz_3=\uu$.
The $\uu$ subproblem is approximated by several gradient steps on the augmented Lagrangian.
The split updates are
\begin{align}
 \zz_1 &\leftarrow \operatorname{soft}_{\lambda_1/\rho_1}(\uu+\yy_1),\\
 \zz_2 &\leftarrow \operatorname{soft}_{\lambda_{\mathrm{TV}}/\rho_2}(D\uu+\yy_2),\\
 \zz_3 &\leftarrow \Pi_{\mathcal B}(\uu+\yy_3).
\end{align}
Scaled dual variables are then updated from the corresponding split residuals.
Amplitude projection is applied during the control update.

Only active splits contribute to the reported residual norms.
After a minimum iteration budget, iterations stop when both primal and dual residuals satisfy the specified absolute and relative tolerances.
We project the returned waveform onto the intersection of the box and band-limit sets by alternating convex projections.
All reported metrics are recomputed from that returned waveform.
This evaluation rule prevents intermediate split variables, which are useful algorithmic auxiliaries but need not satisfy every physical constraint, from entering the reported comparison.
It also distinguishes optimisation progress from the deployable quantity of interest: the fidelity and complexity of one admissible waveform under the declared model and perturbation convention.

\subsection{Reference optimisers and warm starts}

We compare PADMM with GRAPE, the standard \texttt{qucontrol/krotov} implementation, and SciPy L-BFGS-B.
The Krotov baseline uses the monotonic-update framework described in Ref.~\cite{Reich2012}.
For each seed, every method starts from the same Gaussian pulse.
PADMM-Warm first applies a fixed GRAPE budget and then activates the structured updates.
This construction tests whether a short fidelity-oriented stage places the waveform in a more favourable basin.
The shared initialisation makes each seed a paired numerical replicate rather than a method-specific source of variability.
The baselines retain their native update rules, but their returned controls are evaluated under the same model, target, amplitude convention, and post-processing protocol.
Thus, the comparison does not assume that internal iteration counters from heterogeneous solvers are commensurate.

\subsection{Robust perturbations}

We evaluate robustness under detuning, multiplicative amplitude error, and additive linear drift.
The same perturbation function is used for robust training and post-training evaluation.
After scaling or drift is applied, the waveform is clipped to the task-specific amplitude bounds.
Robust training minimises the weighted mean infidelity over the nominal and perturbed scenarios.
The perturbation ensemble is deliberately finite. It is a reproducible stress test defined by detuning, multiplicative scale error, and linear drift, not an experimentally inferred probability distribution.
Robust-training outcomes are therefore interpreted only within the stated ensemble and amplitude-clipping rule.

\subsection{Statistical analysis and reproducibility}

The principal evaluation uses ten fixed seeds, whereas hyperparameter selection uses separate seeds.
The seed is the independent experimental unit, and each method comparison aligns identical seeds.
We report seed-level means and 95\% Student-$t$ intervals.
We assess paired differences with paired $t$ tests and retain Wilcoxon signed-rank values as sensitivity checks.
Effect sizes use the paired standardised difference $d_z$.
Benjamini--Hochberg correction is applied to the prespecified family of primary comparisons.

Perturbation levels are averaged within each seed before robustness intervals are computed.
This procedure prevents perturbation levels generated from one optimised pulse from being treated as independent replicates.
Every result stores the configuration, configuration hash, seed, package versions, metric definitions, and result-schema version.

\section{Results}

\subsection{Nominal fidelity and waveform complexity}

\begin{figure*}
 \centering
 \includegraphics[width=\textwidth]{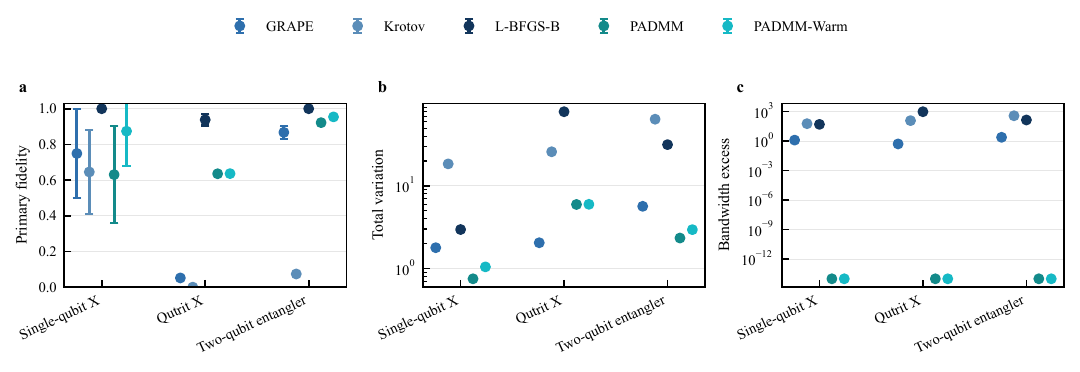}
 \caption{Principal ten-seed benchmark. Points show means, and fidelity error bars show 95\% Student-$t$ intervals across paired seeds. Complexity panels use logarithmic axes; intervals are omitted because symmetric intervals can extend below zero for positive metrics. Qutrit fidelity denotes computational-subspace fidelity. Method colours are fixed across all figures.}
 \label{fig:main}
\end{figure*}

Figure~\ref{fig:main} and Table~\ref{tab:main} report the principal comparison.
The unconstrained quasi-Newton baseline defined the high-fidelity end of the observed numerical trade-off.
PADMM solutions occupied a lower-complexity region, with negligible spectral weight above the configured cutoff.
Warm starting increased fidelity for the single-qubit and two-qubit tasks while retaining the structured waveform character.
L-BFGS-B reached mean fidelities of $1.0000$, $0.9380$, and $1.0000$ for the single-qubit, qutrit, and two-qubit tasks.
PADMM-Warm reached $0.8741$, $0.6363$, and $0.9541$, respectively.
Its mean total variation was $1.043$, $5.977$, and $2.947$, compared with $2.958$, $79.184$, and $31.598$ for L-BFGS-B.

The three panels of Fig.~\ref{fig:main} make this distinction explicit.
Panel (a) establishes the fidelity ordering for each task, whereas panel (b) exposes the associated temporal-variation cost on a logarithmic scale.
Panel (c) provides a separate feasibility diagnostic: the PADMM variants lie close to the numerical floor for bandwidth excess, while the unconstrained solutions generally do not.
Thus, the low-complexity result is not inferred from a single scalar objective.
It is the joint observation of fidelity, temporal variation, and spectral excess under the same seed-paired protocol.
The vertical intervals in panel (a) quantify between-seed uncertainty, but they do not convert the three metrics into a universal ranking.

\begin{table}[t]
\caption{Ten-seed benchmark. Values are means with 95\% seed-level $t$ intervals. Qutrit fidelity is evaluated on the computational subspace.}
\label{tab:main}
\centering
\small
\setlength{\tabcolsep}{3.5pt}
\renewcommand{\arraystretch}{1.08}
\begin{ruledtabular}
\begin{tabular}{@{}llrrrr@{}}
Task & Method & \multicolumn{1}{c}{Fidelity} & \multicolumn{1}{c}{\shortstack{Total\\variation}} & \multicolumn{1}{c}{\shortstack{Bandwidth\\excess}} & \multicolumn{1}{c}{\shortstack{Time\\(s)}}\\
\hline
Single-qubit $X$ & GRAPE & $0.7491$ [0.5012, 0.9971] & $1.781$ & $1.24e+00$ & $1.53$ \\
Single-qubit $X$ & Krotov & $0.6456$ [0.4102, 0.8809] & $18.465$ & $5.82e+01$ & $26.30$ \\
Single-qubit $X$ & L-BFGS-B & $1.0000$ [1.0000, 1.0000] & $2.958$ & $4.95e+01$ & $0.13$ \\
Single-qubit $X$ & PADMM & $0.6306$ [0.3581, 0.9032] & $0.748$ & $2.27e-30$ & $5.30$ \\
Single-qubit $X$ & PADMM-Warm & $0.8741$ [0.6794, 1.0000] & $1.043$ & $2.53e-30$ & $5.58$ \\
Qutrit $X$ & GRAPE & $0.0517$ [0.0481, 0.0554] & $2.045$ & $5.09e-01$ & $3.52$ \\
Qutrit $X$ & Krotov & $0.0001$ [0.0000, 0.0001] & $25.897$ & $1.19e+02$ & $59.42$ \\
Qutrit $X$ & L-BFGS-B & $0.9380$ [0.9061, 0.9699] & $79.184$ & $9.54e+02$ & $1.55$ \\
Qutrit $X$ & PADMM & $0.6355$ [0.6354, 0.6356] & $5.952$ & $2.68e-29$ & $50.62$ \\
Qutrit $X$ & PADMM-Warm & $0.6363$ [0.6362, 0.6364] & $5.977$ & $2.23e-29$ & $48.20$ \\
Two-qubit entangler & GRAPE & $0.8675$ [0.8323, 0.9026] & $5.654$ & $2.39e+00$ & $5.93$ \\
Two-qubit entangler & Krotov & $0.0738$ [0.0736, 0.0740] & $64.259$ & $3.72e+02$ & $81.15$ \\
Two-qubit entangler & L-BFGS-B & $1.0000$ [1.0000, 1.0000] & $31.598$ & $1.38e+02$ & $3.65$ \\
Two-qubit entangler & PADMM & $0.9219$ [0.9111, 0.9327] & $2.333$ & $3.12e-30$ & $22.54$ \\
Two-qubit entangler & PADMM-Warm & $0.9541$ [0.9522, 0.9560] & $2.947$ & $4.53e-30$ & $30.48$ \\
\end{tabular}
\end{ruledtabular}
\end{table}

\subsection{Fidelity--complexity trade-off}

\begin{figure*}
 \centering
 \includegraphics[width=\textwidth]{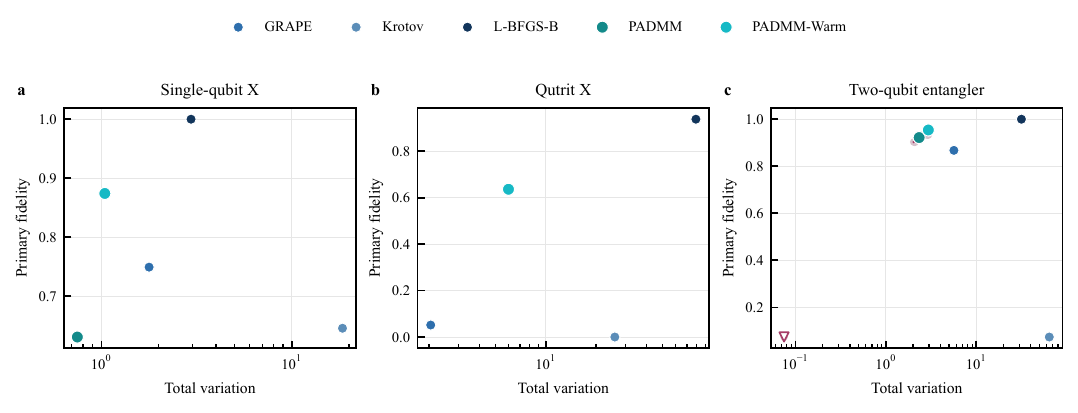}
 \caption{Fidelity--complexity trade-off across the three tasks. Each principal-comparison point is a ten-seed mean, and pale points show the supporting PADMM parameter scan. The horizontal axis is logarithmic. A zero total-variation value, when present in the scan, is plotted at one tenth of the smallest positive value and labelled ``0 TV''; the unmodified value is retained in the source-data file. Structured solutions occupy the upper-left region, subject to the task-specific fidelity requirement.}
 \label{fig:tradeoff}
\end{figure*}

Figure~\ref{fig:tradeoff} shows that the methods do not admit a single performance ranking.
L-BFGS-B prioritises nominal fidelity, whereas PADMM and PADMM-Warm reduce temporal and spectral complexity.
The appropriate comparison is therefore a task-specific frontier, not a claim of universal superiority.
In the two-qubit task, PADMM-Warm reduced total variation by a factor of approximately $10.7$ relative to L-BFGS-B while retaining a mean fidelity of $0.9541$.

Figure~\ref{fig:tradeoff} resolves the same data as task-specific frontiers rather than method bars.
Within each panel, a displacement towards the upper left represents higher primary fidelity at lower total variation.
The dark method markers are the principal ten-seed means used for the manuscript's quantitative comparisons.
By contrast, the pale points are a supporting PADMM parameter scan and therefore show reachable design alternatives, not additional independent replicates.
The scan is useful because it reveals that the structured formulation can move along the trade-off surface as its penalties change.
It is not evidence that every pale point would retain its location under a new seed ensemble.
The display convention for zero total variation is solely required by the logarithmic axis; the unmodified value remains available in the source data.

\subsection{Qutrit leakage changes the interpretation}

\begin{figure*}
 \centering
 \includegraphics[width=\textwidth]{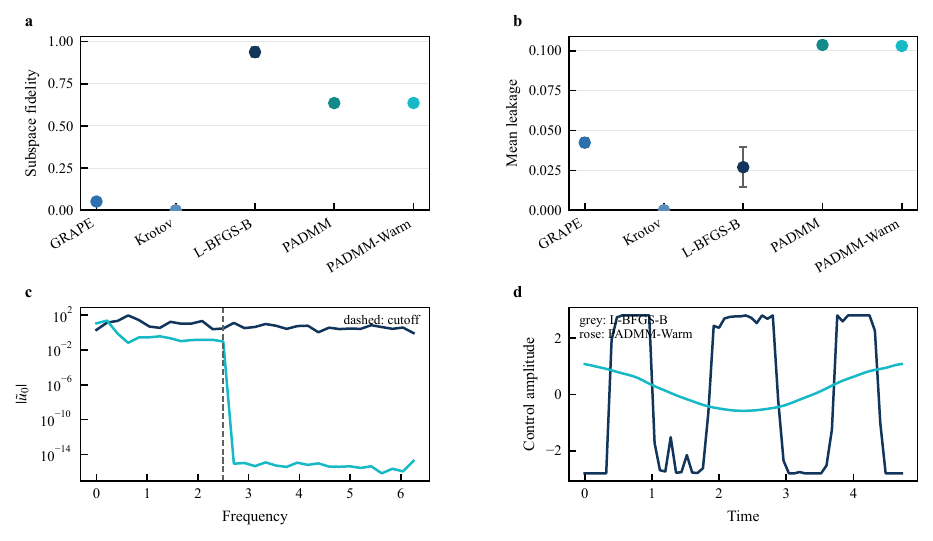}
 \caption{Qutrit performance and representative pulse structure. (a) Computational-subspace fidelity. (b) Mean leakage. (c) First control-channel spectra for representative L-BFGS-B and PADMM-Warm pulses; the dashed line marks the PADMM cutoff. (d) Corresponding first-channel waveforms. Error bars in (a,b) are 95\% Student-$t$ intervals across ten seeds. The waveforms are illustrative; all quantitative comparisons use all seeds.}
 \label{fig:qutrit}
\end{figure*}

The qutrit metrics in Table~\ref{tab:qutrit} distinguish logical performance from leakage.
Full-space fidelity alone does not capture this distinction.
The pulse and spectrum comparison in Fig.~\ref{fig:qutrit} illustrates the structural difference between an unconstrained high-fidelity pulse and a band-limited PADMM-Warm pulse.
PADMM-Warm produced a mean subspace fidelity of $0.6363$ and mean leakage of $0.1029$.
L-BFGS-B reached $0.9380$ subspace fidelity with mean leakage of $0.0270$, but its total variation was more than an order of magnitude larger. Thus, the structured solutions reduce waveform complexity at a substantial cost in qutrit fidelity and leakage.

Figure~\ref{fig:qutrit} shows why this cost cannot be diagnosed from a waveform alone.
Panels (a) and (b) place subspace fidelity and mean leakage side by side, so the reader can distinguish a logical-control deficit from population loss out of the computational subspace.
Panels (c) and (d) then provide a representative mechanism-level view for the same comparison: the PADMM-Warm spectrum is sharply suppressed beyond the dashed cutoff, and its displayed control is correspondingly smoother than the L-BFGS-B trace.
Those lower panels illustrate consistency with the aggregate bandwidth and total-variation metrics; they do not provide an additional statistical comparison because they show one selected stored control per method.
Taken together, the four panels show that spectral admissibility was obtained by the structured search, but not without a measurable qutrit-performance penalty.

\begin{table*}
\caption{Qutrit metrics across ten seeds. Intervals are seed-level 95\% $t$ intervals.}
\label{tab:qutrit}
\begin{ruledtabular}
\begin{tabular}{lccc}
Method & Subspace fidelity & Full-space fidelity & Mean leakage\\
\hline
GRAPE & $0.0517$ & $0.1416$ & $0.0424$ \\
Krotov & $0.0001$ & $0.1111$ & $0.0001$ \\
L-BFGS-B & $0.9380$ & $0.9401$ & $0.0270$ \\
PADMM & $0.6355$ & $0.6663$ & $0.1036$ \\
PADMM-Warm & $0.6363$ & $0.6672$ & $0.1029$ \\
\end{tabular}
\end{ruledtabular}
\end{table*}

\subsection{Paired comparisons and robustness}

The paired analysis in Table~\ref{tab:paired} uses the correlation induced by common initial seeds.
It replaces the independent-sample analysis used in the earlier project.
The robustness analysis averages perturbation levels within each optimised pulse before aggregating over seeds.
Warm starting increased PADMM fidelity by $0.2435$ for the single-qubit task, $0.0008$ for the qutrit, and $0.0322$ for the two-qubit task.
The corresponding paired $t$ tests remained significant after false-discovery-rate correction. These tests quantify the benefit of warm starting within PADMM; they do not establish an advantage over L-BFGS-B.

\begin{figure*}
 \centering
 \includegraphics[width=\textwidth]{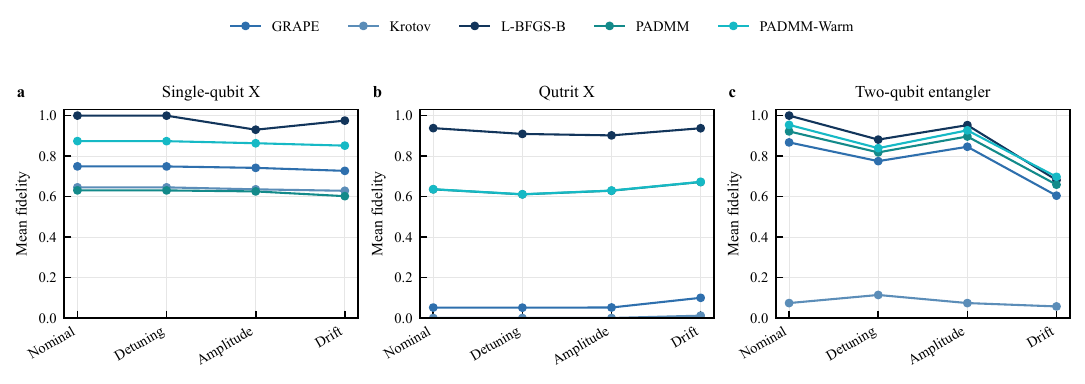}
 \caption{Post-training robustness under detuning, multiplicative amplitude error, and additive control drift. Each point is first averaged over the specified perturbation levels within a seed and then averaged across ten seeds ($n=10$). All panels use the same 0--1.03 fidelity range. Lines connect conditions only to aid reading and do not imply a continuous perturbation sweep.}
 \label{fig:robustness}
\end{figure*}

Figure~\ref{fig:robustness} provides the corresponding perturbation-level view.
All three task panels use the same fidelity range, which makes the pronounced drift-related decrease in the two-qubit task visually comparable with the more modest changes in the single-qubit panel.
Each curve point is already an average over the specified perturbation levels within a seed before the ten-seed aggregation, so neighbouring markers on a line are not treated as independent experimental replicates.
The connecting lines aid comparison among the four named conditions only; they should not be read as a continuous response function.
This distinction is important because robustness in the present study is an evaluation under a finite, prespecified perturbation set rather than a certification over all hardware errors.

\begin{table}[t]
\caption{Prespecified paired comparisons on matching seeds. The effect size is Cohen's $d_z$; $q$ values use Benjamini--Hochberg correction over the displayed family.}
\label{tab:paired}
\centering
\scriptsize
\setlength{\tabcolsep}{2.5pt}
\renewcommand{\arraystretch}{1.10}
\begin{ruledtabular}
\begin{tabular}{@{}llrrrrr@{}}
Task & Comparison & \multicolumn{1}{c}{\shortstack{Mean\\difference}} & \multicolumn{1}{c}{95\% CI} & $d_z$ & $p$ & $q$\\
\hline
Single-qubit $X$ & \shortstack[l]{PADMM-Warm\\vs.\ PADMM} & $0.2435$ & [0.0524, 0.4346] & $0.91$ & $0.0181$ & $0.0226$ \\
Single-qubit $X$ & \shortstack[l]{PADMM\\vs.\ L-BFGS-B} & $-0.3694$ & [-0.6419, -0.0968] & $-0.97$ & $0.0135$ & $0.0183$ \\
Single-qubit $X$ & \shortstack[l]{PADMM-Warm\\vs.\ L-BFGS-B} & $-0.1259$ & [-0.3206, 0.0688] & $-0.46$ & $0.178$ & $0.178$ \\
Single-qubit $X$ & \shortstack[l]{PADMM\\vs.\ GRAPE} & $-0.1185$ & [-0.2388, 0.0018] & $-0.70$ & $0.0528$ & $0.0566$ \\
Single-qubit $X$ & \shortstack[l]{PADMM-Warm\\vs.\ GRAPE} & $0.1250$ & [0.0089, 0.2410] & $0.77$ & $0.0376$ & $0.0434$ \\
Qutrit $X$ & \shortstack[l]{PADMM-Warm\\vs.\ PADMM} & $0.0008$ & [0.0008, 0.0008] & $88.19$ & $4.99e-19$ & $2.49e-18$ \\
Qutrit $X$ & \shortstack[l]{PADMM\\vs.\ L-BFGS-B} & $-0.3025$ & [-0.3344, -0.2706] & $-6.78$ & $4.94e-09$ & $1.27e-08$ \\
Qutrit $X$ & \shortstack[l]{PADMM-Warm\\vs.\ L-BFGS-B} & $-0.3017$ & [-0.3336, -0.2697] & $-6.76$ & $5.06e-09$ & $1.27e-08$ \\
Qutrit $X$ & \shortstack[l]{PADMM\\vs.\ GRAPE} & $0.5838$ & [0.5802, 0.5874] & $115.29$ & $4.47e-20$ & $3.36e-19$ \\
Qutrit $X$ & \shortstack[l]{PADMM-Warm\\vs.\ GRAPE} & $0.5846$ & [0.5810, 0.5882] & $115.43$ & $4.42e-20$ & $3.36e-19$ \\
Two-qubit entangler & \shortstack[l]{PADMM-Warm\\vs.\ PADMM} & $0.0322$ & [0.0225, 0.0419] & $2.37$ & $3.75e-05$ & $7.03e-05$ \\
Two-qubit entangler & \shortstack[l]{PADMM\\vs.\ L-BFGS-B} & $-0.0781$ & [-0.0889, -0.0673] & $-5.17$ & $5.36e-08$ & $1.15e-07$ \\
Two-qubit entangler & \shortstack[l]{PADMM-Warm\\vs.\ L-BFGS-B} & $-0.0459$ & [-0.0477, -0.0440] & $-17.49$ & $1.04e-12$ & $3.89e-12$ \\
Two-qubit entangler & \shortstack[l]{PADMM\\vs.\ GRAPE} & $0.0545$ & [0.0254, 0.0835] & $1.34$ & $0.00218$ & $0.00326$ \\
Two-qubit entangler & \shortstack[l]{PADMM-Warm\\vs.\ GRAPE} & $0.0867$ & [0.0532, 0.1201] & $1.85$ & $0.000242$ & $0.000404$ \\
\end{tabular}
\end{ruledtabular}
\end{table}

\begin{table*}
\caption{Post-training robustness. Perturbation levels are first averaged within each seed; entries then report the mean across seeds.}
\label{tab:robustness}
\begin{ruledtabular}
\begin{tabular}{llcccc}
Task & Method & Nominal & Detuning & Amplitude error & Control drift\\
\hline
Single-qubit $X$ & GRAPE & $0.7491$ & $0.7487$ & $0.7417$ & $0.7269$ \\
Single-qubit $X$ & Krotov & $0.6456$ & $0.6454$ & $0.6357$ & $0.6285$ \\
Single-qubit $X$ & L-BFGS-B & $1.0000$ & $0.9995$ & $0.9302$ & $0.9752$ \\
Single-qubit $X$ & PADMM & $0.6306$ & $0.6303$ & $0.6253$ & $0.6019$ \\
Single-qubit $X$ & PADMM-Warm & $0.8741$ & $0.8736$ & $0.8634$ & $0.8515$ \\
Qutrit $X$ & GRAPE & $0.0517$ & $0.0511$ & $0.0520$ & $0.0998$ \\
Qutrit $X$ & Krotov & $0.0001$ & $0.0001$ & $0.0001$ & $0.0117$ \\
Qutrit $X$ & L-BFGS-B & $0.9380$ & $0.9093$ & $0.9021$ & $0.9375$ \\
Qutrit $X$ & PADMM & $0.6355$ & $0.6107$ & $0.6288$ & $0.6716$ \\
Qutrit $X$ & PADMM-Warm & $0.6363$ & $0.6114$ & $0.6296$ & $0.6728$ \\
Two-qubit entangler & GRAPE & $0.8675$ & $0.7749$ & $0.8460$ & $0.6044$ \\
Two-qubit entangler & Krotov & $0.0738$ & $0.1139$ & $0.0738$ & $0.0572$ \\
Two-qubit entangler & L-BFGS-B & $1.0000$ & $0.8813$ & $0.9529$ & $0.6816$ \\
Two-qubit entangler & PADMM & $0.9219$ & $0.8182$ & $0.8968$ & $0.6586$ \\
Two-qubit entangler & PADMM-Warm & $0.9541$ & $0.8391$ & $0.9272$ & $0.6976$ \\
\end{tabular}
\end{ruledtabular}
\end{table*}

\subsection{Ablations and robust training}

The supporting studies use three seeds that are disjoint from the principal evaluation.
Constraint ablations identify which split terms shape the observed fidelity--complexity compromise.
The robust-training comparison tests whether optimisation over a finite perturbation ensemble improves drift performance beyond the structured warm-started pulse.
Both robust-training variants use 50 GRAPE warm-start iterations, 120 PADMM outer iterations, and five inner gradient steps.
These studies are exploratory because their seed count is smaller than that of the principal benchmark.
Robust training increased mean drift fidelity from $0.2109$ to $0.2161$ for the qutrit and from $0.6248$ to $0.6617$ for the two-qubit task.
For the two-qubit task, this gain accompanied a decrease in nominal fidelity from $0.9002$ to $0.8959$.

Table~\ref{tab:supporting} separates two questions that the principal comparison cannot answer alone.
The ablations test whether the observed structured solution depends on sparsity, temporal-variation, and bandwidth components of the split objective.
The robust-training rows test a different proposition: whether exposing the warm-started control to the finite perturbation ensemble changes drift performance.
The qutrit improvement is small on the reported scale, whereas the two-qubit improvement is larger but accompanied by a nominal-fidelity reduction.
Robustness is therefore not an unconditional gain; it reallocates performance across nominal and perturbed conditions.
Because these studies use only three disjoint seeds, they guide interpretation and future hyperparameter selection but do not replace the ten-seed principal comparison.

\begin{table*}
\caption{Supporting three-seed studies using seeds disjoint from the principal evaluation. The ablation entries report primary fidelity; robust-training entries report fidelity under control drift.}
\label{tab:supporting}
\begin{ruledtabular}
\begin{tabular}{llcc}
Task & Study variant & Mean & 95\% seed-level CI\\
\hline
Qutrit $X$ & Ablation: full & $0.6355$ & [0.6353, 0.6357] \\
Qutrit $X$ & Ablation: no sparsity & $0.6374$ & [0.6372, 0.6376] \\
Qutrit $X$ & Ablation: no tv & $0.6354$ & [0.6352, 0.6356] \\
Qutrit $X$ & Ablation: no bandlimit & $0.6370$ & [0.6369, 0.6372] \\
Qutrit $X$ & Ablation: fidelity only & $0.6105$ & [0.6023, 0.6186] \\
Two-qubit entangler & Ablation: full & $0.9004$ & [0.8656, 0.9353] \\
Two-qubit entangler & Ablation: no sparsity & $0.9065$ & [0.8352, 0.9778] \\
Two-qubit entangler & Ablation: no tv & $0.9039$ & [0.8755, 0.9324] \\
Two-qubit entangler & Ablation: no bandlimit & $0.9234$ & [0.9083, 0.9386] \\
Two-qubit entangler & Ablation: fidelity only & $0.1466$ & [-0.0009, 0.2942] \\
Qutrit $X$ & Robust training: PADMM-Warm & $0.2109$ & [0.1809, 0.2409] \\
Qutrit $X$ & Robust training: PADMM-Warm-Robust & $0.2161$ & [0.1863, 0.2458] \\
Two-qubit entangler & Robust training: PADMM-Warm & $0.6248$ & [0.5243, 0.7253] \\
Two-qubit entangler & Robust training: PADMM-Warm-Robust & $0.6617$ & [0.5722, 0.7513] \\
\end{tabular}
\end{ruledtabular}
\end{table*}

\subsection{Pulse morphology exposes the structured search outcome}

Figure~\ref{fig:morphology} compares representative, matching-seed controls from L-BFGS-B and PADMM-Warm.
For the single-qubit and qutrit tasks, the two quadratures are represented as envelope amplitude and unwrapped phase.
For the two-qubit task, the four local control channels are shown directly.
The PADMM-Warm traces are smoother and use smaller channel excursions in the displayed representatives.
This visualisation links the total-variation metric to time-domain waveform structure; it is not an additional method ranking.

\begin{center}
 \centering
 \includegraphics[width=0.90\textwidth]{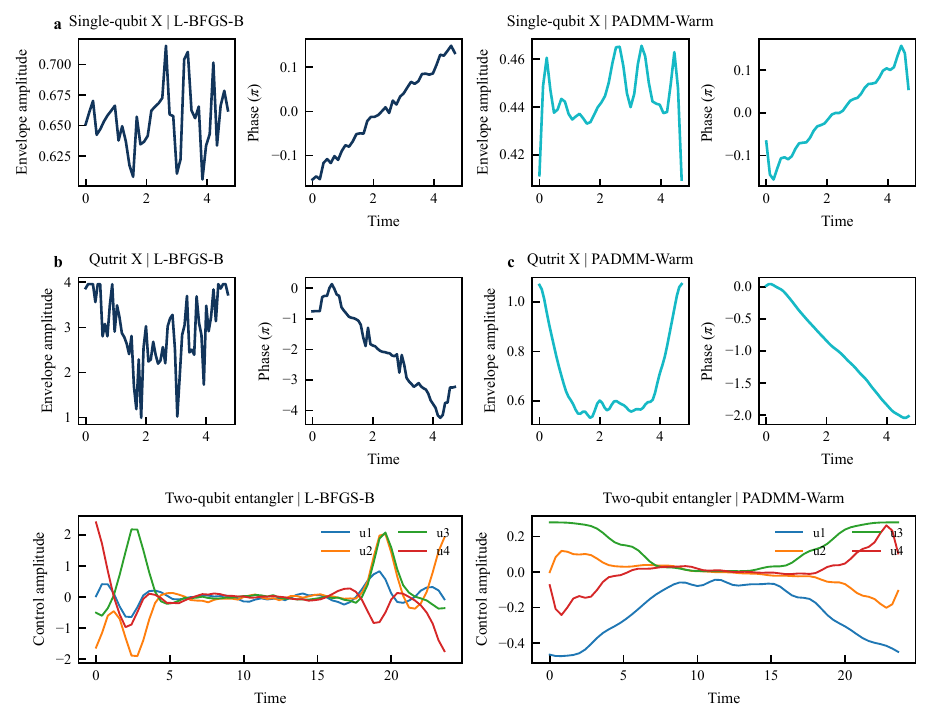}\par\smallskip
 \refstepcounter{figure}\textbf{FIG.~\thefigure.}\ Representative pulse morphology for matching seed 29. Rows show the single-qubit, qutrit, and two-qubit tasks. The first two rows reconstruct envelope amplitude and unwrapped phase from the two quadrature controls. The last row shows the four local two-qubit control channels. Colours identify methods: navy for L-BFGS-B and cyan for PADMM-Warm. These representative traces visualise the waveform structure; the quantitative conclusions remain based on all ten seeds.\label{fig:morphology}
\end{center}

The row structure in Fig.~\ref{fig:morphology} links the aggregate trade-off to concrete pulse forms.
For the single-qubit and qutrit tasks, separating envelope amplitude from unwrapped phase makes it possible to identify whether apparent smoothness originates from the amplitude profile, phase progression, or both.
For the two-qubit task, the local channels remain separate because combining them into one envelope would obscure the channel-wise excursions that contribute to the control burden.
Across the displayed seed-29 controls, the cyan PADMM-Warm traces vary more gradually than the navy L-BFGS-B traces, in agreement with the lower total-variation values in Table~\ref{tab:main} and Fig.~\ref{fig:tradeoff}.
This agreement is descriptive rather than inferential: the representative panels help interpret the aggregate metric, whereas the ten-seed tables remain the evidence for the reported method comparison.

\subsection{Two-parameter robustness maps}

Figure~\ref{fig:maps} evaluates the same representative stored controls over a fixed grid of relative detuning and multiplicative amplitude scale.
Each grid point is propagated afresh, with amplitude clipping applied after scaling under the same convention used for the robustness evaluation.
The qutrit panels report computational-subspace fidelity, whereas the two-qubit panels report full-space fidelity.
The maps show task- and method-specific robustness regions rather than a universal structured-control advantage.

Figure~\ref{fig:maps} adds information that is not visible in the one-dimensional robustness curves.
The white star identifies the nominal control condition, while departures along the horizontal and vertical directions separately alter detuning and amplitude scale.
The common colour scale and repeated contour levels allow the shape and extent of high-fidelity regions to be compared across methods and tasks without a panel-specific rescaling.
For example, a broad dark region indicates tolerance within this two-parameter grid, whereas a narrow or displaced dark region indicates sensitivity to at least one of the two perturbations.
These panels are deterministic re-evaluations of one stored seed-29 pulse per method, not confidence maps across the ten-seed population.
They therefore complement, but do not replace, the seed-level robustness statistics in Table~\ref{tab:robustness} and Fig.~\ref{fig:robustness}.

\begin{center}
 \centering
 \includegraphics[width=0.90\textwidth]{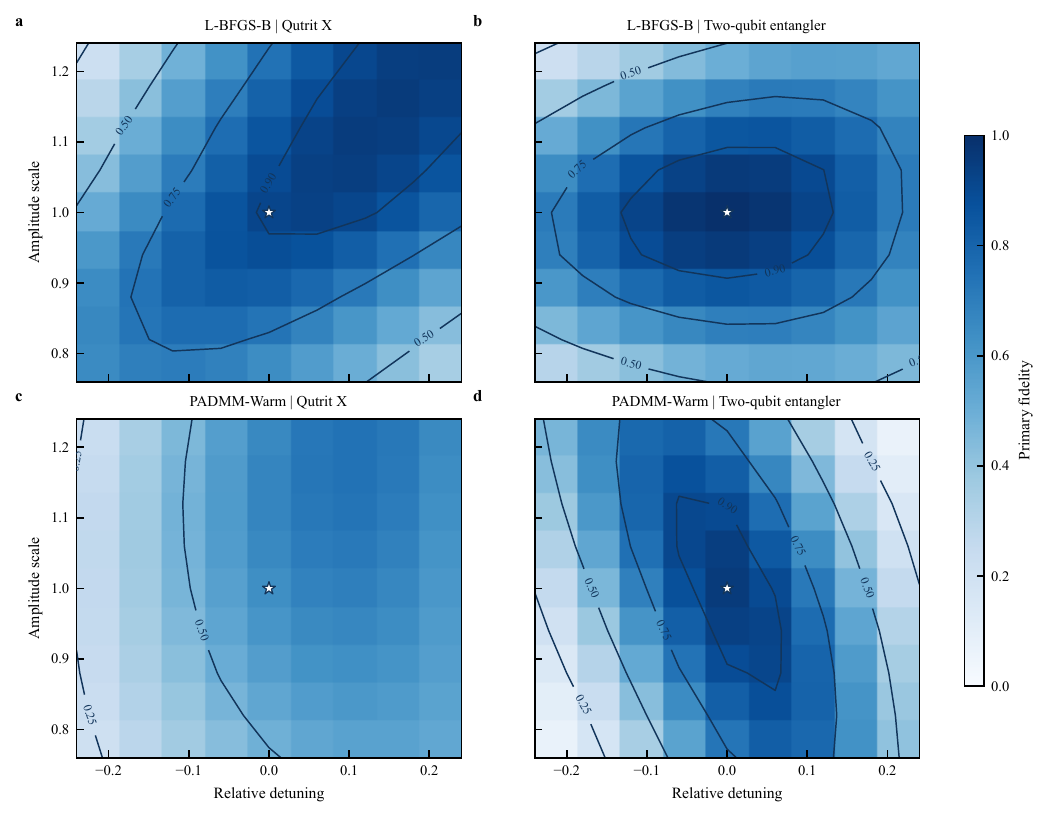}\par\smallskip
 \refstepcounter{figure}\textbf{FIG.~\thefigure.}\ Two-parameter robustness maps for representative seed-29 L-BFGS-B and PADMM-Warm controls. Columns show the qutrit and two-qubit tasks; rows show methods. Each pixel is a fresh propagation over relative detuning from $-0.24$ to $0.24$ and amplitude scale from $0.76$ to $1.24$. The common blue colour scale spans primary fidelity from 0 to 1, and contours mark 0.25, 0.50, 0.75, and 0.90. Stars mark the nominal point. Qutrit panels use computational-subspace fidelity; two-qubit panels use full-space fidelity.\label{fig:maps}
\end{center}

\subsection{Single-qubit state trajectories}

Figure~\ref{fig:bloch} maps three representative single-qubit controls to their state trajectories from $|0\rangle$.
All trajectories share the same initial state, target gate, axis limits, and viewing angle.
The plot provides a dynamical interpretation of pulse shape, but applies only to the single-qubit benchmark and does not establish a general performance ordering.

The three panels in Fig.~\ref{fig:bloch} use identical axes and viewing angles so that path geometry can be compared without a camera-dependent visual effect.
The navy marker fixes the shared initial state, and the cyan marker records the final state reached by each stored control under the same propagation model used for the benchmark.
The trajectories therefore provide a compact dynamical check on how distinct control waveforms traverse the Bloch sphere, rather than another fidelity summary.
In particular, the figure should not be used to extrapolate to qutrit leakage or two-qubit entangling dynamics, which require different state spaces and metrics.

\begin{center}
 \centering
 \includegraphics[width=\textwidth]{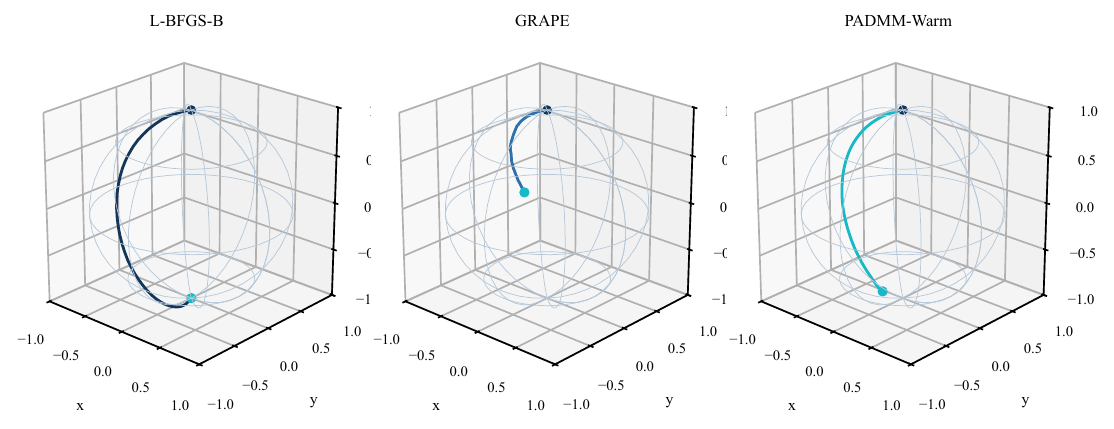}\par\smallskip
 \refstepcounter{figure}\textbf{FIG.~\thefigure.}\ Bloch-sphere trajectories for representative seed-29 single-qubit controls. The state begins at $|0\rangle$ (navy marker) and evolves under L-BFGS-B, GRAPE, or PADMM-Warm pulses to the final state (cyan marker). All panels use identical axes and viewing angles. This single-qubit visualisation links pulse morphology to state evolution; it is not used as evidence of a general method advantage.\label{fig:bloch}
\end{center}

\section{Discussion}

The corrected experiments support a bounded conclusion.
The inexact PADMM framework provides direct control over pulse bandwidth and temporal structure.
It neither consistently maximises nominal fidelity nor minimises wall-clock time.
Its practical value is the explicit construction of low-complexity solutions and a transparent view of the fidelity--complexity frontier.
The aggregate panels, representative waveform diagnostics, two-parameter maps, and Bloch trajectories are deliberately complementary: together they relate numerical trade-offs to pulse and state structure, while preserving the distinction between ten-seed evidence and representative visualisation.

The revised two-qubit task is more informative than the original benchmark.
Removing a direct control proportional to the target generator prevents the optimiser from implementing the entangler through one trivial channel.
High fidelity remains possible through local rotations and the fixed interaction, but the search is more demanding.

From an implementation perspective, the value of this formulation is not that one penalty setting yields the best control for every task.
Instead, it makes the compromise visible in quantities that connect directly to pulse engineering: amplitude admissibility, temporal variation, spectral support, nominal gate quality, and response to the declared perturbations.
The resulting frontier can guide selection when a pulse must satisfy more than one operational requirement.
That selection still requires task-dependent thresholds, since the present simulations do not establish a device-independent conversion from total variation or bandwidth excess to experimental error.

The qutrit analysis also illustrates why a single trace fidelity can be insufficient.
Computational-subspace fidelity and leakage answer different questions.
A waveform may approximate the desired full-space unitary while still entering a region that is unsuitable for logical control.
Reporting both metrics exposes this limitation.

The same separation applies to structural and dynamical diagnostics.
A smooth representative waveform or a broad region in a two-parameter map can make the origin of a trade-off easier to inspect, but it does not supersede the seed-level estimates used for the main claims.
Conversely, an aggregate improvement in one complexity metric does not establish robustness against errors absent from the perturbation ensemble.
Keeping these evidence layers separate is central to a fair methods comparison.

Several limitations remain.
The models are closed systems and exclude decoherence and calibration dynamics.
The Fourier cutoff is an ideal discrete projection rather than a transfer function measured from hardware.
The inexact control update does not inherit standard convex ADMM convergence guarantees because gate synthesis is nonconvex.
Ten seeds support paired numerical comparisons, but they cannot establish broad hardware generality.

Computational-cost comparisons also require care.
Krotov uses a native sequential update, whereas the other methods expose explicit objective-gradient calls.
We therefore emphasise wall-clock time under one software environment and avoid treating noncommensurate internal counters as equivalent.

Future work should test the structured formulation with open-system dynamics and measured hardware transfer functions.
An adaptive penalty strategy may also reduce sensitivity to fixed PADMM hyperparameters.
These extensions should preserve seed-level pairing and task-aware physical metrics.

\section{Conclusion}

We developed and audited an inexact proximal-ADMM framework for quantum pulses subject to amplitude, bandwidth, sparsity, and temporal-variation constraints.
The revised benchmark removes a direct two-qubit control shortcut and reports qutrit logical fidelity alongside leakage.
Across the simulated tasks, PADMM identifies low-complexity pulses but does not replace unconstrained high-fidelity optimisation.
The method is therefore best viewed as a constraint-native tool for exploring implementability trade-offs.

\clearpage
\bibliography{references_qoc}

@article{Khaneja2005,
  author = {Khaneja, Navin and Reiss, Timo and Kehlet, Cindie and Schulte-Herbr{\"u}ggen, Thomas and Glaser, Steffen J.},
  title = {Optimal control of coupled spin dynamics: Design of {NMR} pulse sequences by gradient ascent algorithms},
  journal = {Journal of Magnetic Resonance},
  volume = {172},
  number = {2},
  pages = {296--305},
  year = {2005}
}

@article{Palao2003,
  author = {Palao, Jos{\'e} P. and Kosloff, Ronnie},
  title = {Optimal control theory for unitary transformations},
  journal = {Physical Review A},
  volume = {68},
  pages = {062308},
  year = {2003}
}

@article{Goerz2019,
  author = {Goerz, Michael H. and Basilewitsch, Daniel and Gago-Encinas, Fernando and Egger, Daniel J. and Koch, Christiane P. and Whaley, K. Birgitta},
  title = {Krotov: A {Python} implementation of {Krotov's} method for quantum optimal control},
  journal = {SciPost Physics},
  volume = {7},
  number = {6},
  pages = {080},
  year = {2019}
}

@article{Boyd2011,
  author = {Boyd, Stephen and Parikh, Neal and Chu, Eric and Peleato, Borja and Eckstein, Jonathan},
  title = {Distributed optimization and statistical learning via the alternating direction method of multipliers},
  journal = {Foundations and Trends in Machine Learning},
  volume = {3},
  number = {1},
  pages = {1--122},
  year = {2011}
}

@article{Parikh2014,
  author = {Parikh, Neal and Boyd, Stephen},
  title = {Proximal algorithms},
  journal = {Foundations and Trends in Optimization},
  volume = {1},
  number = {3},
  pages = {127--239},
  year = {2014}
}

@article{Eitan2011,
  author = {Eitan, Reuven and Mundt, Michael and Tannor, David J.},
  title = {Optimal control with accelerated convergence: Combining the {Krotov} and quasi-{Newton} methods},
  journal = {Physical Review A},
  volume = {83},
  pages = {053426},
  year = {2011}
}

@article{VanFrank2016,
  author = {van Frank, Steffen and Wilhelm-Mauch, Florian and Schwarzkopf, Andreas and Sch{\"a}ff, Julius F. and Braukmann, Daniel and Schweigler, Thomas and H{\"a}ffner, Hartmut and Schmiedmayer, J{\"o}rg},
  title = {Optimal control of complex atomic quantum systems},
  journal = {Scientific Reports},
  volume = {6},
  pages = {34187},
  year = {2016}
}

@article{Koch2016,
  author = {Koch, Christiane P.},
  title = {Controlling open quantum systems: Tools, achievements, and limitations},
  journal = {Journal of Physics: Condensed Matter},
  volume = {28},
  number = {21},
  pages = {213001},
  year = {2016}
}

@article{Rabitz2000,
  author = {Rabitz, Herschel and de Vivie-Riedle, Regina and Motzkus, Marcus and Kompa, Karl},
  title = {Whither the future of controlling quantum phenomena?},
  journal = {Science},
  volume = {288},
  number = {5467},
  pages = {824--828},
  year = {2000}
}

@article{Werschnik2007,
  author = {Werschnik, Jan and Gross, E. K. U.},
  title = {Quantum optimal control theory},
  journal = {Journal of Physics B: Atomic, Molecular and Optical Physics},
  volume = {40},
  number = {18},
  pages = {R175--R211},
  year = {2007}
}

@article{Brif2010,
  author = {Brif, Constantin and Chakrabarti, Raj and Rabitz, Herschel},
  title = {Control of quantum phenomena: Past, present and future},
  journal = {New Journal of Physics},
  volume = {12},
  pages = {075008},
  year = {2010}
}

@article{Glaser2015,
  author = {Glaser, Steffen J. and Boscain, Ugo and Calarco, Tommaso and Koch, Christiane P. and K{\"o}ckenberger, Wilfred and Kosloff, Ronnie and Kuprov, Ilya and Luy, Burkhard and Schirmer, Sophie and Schulte-Herbr{\"u}ggen, Thomas and Sugny, Dominique and Wilhelm, Frank K.},
  title = {Training Schr{\"o}dinger's cat: Quantum optimal control},
  journal = {The European Physical Journal D},
  volume = {69},
  pages = {279},
  year = {2015}
}

@book{DAlessandro2008,
  author = {D'Alessandro, Domenico},
  title = {Introduction to Quantum Control and Dynamics},
  publisher = {Chapman and Hall/CRC},
  year = {2008}
}

@article{Caneva2011,
  author = {Caneva, Tommaso and Calarco, Tommaso and Montangero, Simone},
  title = {Chopped random-basis quantum optimization},
  journal = {Physical Review A},
  volume = {84},
  pages = {022326},
  year = {2011}
}

@article{Doria2011,
  author = {Doria, Paolo and Calarco, Tommaso and Montangero, Simone},
  title = {Optimal control technique for many-body quantum dynamics},
  journal = {Physical Review Letters},
  volume = {106},
  pages = {190501},
  year = {2011}
}

@article{Machnes2011,
  author = {Machnes, Shai and Sander, Ulf and Glaser, Steffen J. and de Fouqui{\`e}res, Pierre and Gruslys, Audrius and Schirmer, Sophie and Schulte-Herbr{\"u}ggen, Thomas},
  title = {Comparing, optimizing, and benchmarking quantum-control algorithms in a unifying programming framework},
  journal = {Physical Review A},
  volume = {84},
  pages = {022305},
  year = {2011}
}

@article{deFouquieres2011,
  author = {de Fouqui{\`e}res, Pierre and Schirmer, Sophie and Glaser, Steffen J. and Gruslys, Audrius},
  title = {Second order gradient ascent pulse engineering},
  journal = {Journal of Magnetic Resonance},
  volume = {212},
  number = {2},
  pages = {412--417},
  year = {2011}
}

@article{Machnes2018,
  author = {Machnes, Shai and Tannor, David J. and Wilhelm, Frank K. and Katz, N. and Lechner, Wolfgang},
  title = {Tunable, flexible, and efficient optimization of control pulses for practical qubits},
  journal = {Physical Review Letters},
  volume = {120},
  pages = {150401},
  year = {2018}
}

@article{Motzoi2009,
  author = {Motzoi, Felix and Gambetta, Jay M. and Rebentrost, Patrick and Wilhelm, Frank K.},
  title = {Simple pulses for elimination of leakage in weakly nonlinear qubits},
  journal = {Physical Review Letters},
  volume = {103},
  pages = {110501},
  year = {2009}
}

@article{Schutjens2013,
  author = {Schutjens, Robin and Dagga, Faris A. and Egger, Daniel J. and Wilhelm, Frank K.},
  title = {Single-qubit gates in frequency-crowded transmon systems},
  journal = {Physical Review A},
  volume = {88},
  pages = {052330},
  year = {2013}
}

@article{Goerz2015,
  author = {Goerz, Michael H. and Reich, Daniel M. and Koch, Christiane P.},
  title = {Hybrid optimization schemes for quantum control},
  journal = {EPJ Quantum Technology},
  volume = {2},
  pages = {21},
  year = {2015}
}

@article{Egger2014,
  author = {Egger, Daniel J. and Wilhelm, Frank K.},
  title = {Adaptive hybrid optimal quantum control for imprecisely characterized systems},
  journal = {Physical Review Letters},
  volume = {112},
  pages = {240503},
  year = {2014}
}

@article{Kelly2014,
  author = {Kelly, Julian and Barends, Rami and Fowler, Austin G. and Megrant, Anthony and Jeffrey, Evan and White, Theodore C. and Sank, Daniel and Mutus, Josh Y. and Campbell, B. and Chen, Yu and Chen, Zijun and Chiaro, Ben and Dunsworth, Andrew and Neill, Charles and O'Malley, Peter J. J. and Roushan, Pedram and Vainsencher, Ami and Wenner, J. and Korotkov, Alexander N. and Cleland, Andrew N. and Martinis, John M.},
  title = {Optimal quantum control using randomized benchmarking},
  journal = {Physical Review Letters},
  volume = {112},
  pages = {240504},
  year = {2014}
}

@article{Chakrabarti2007,
  author = {Chakrabarti, Raj and Rabitz, Herschel},
  title = {Quantum control landscapes},
  journal = {International Reviews in Physical Chemistry},
  volume = {26},
  number = {4},
  pages = {671--735},
  year = {2007}
}

@book{Nocedal2006,
  author = {Nocedal, Jorge and Wright, Stephen J.},
  title = {Numerical Optimization},
  publisher = {Springer},
  edition = {2},
  year = {2006}
}

@article{Reich2012,
  author = {Reich, Daniel M. and Ndong, Moussa and Koch, Christiane P.},
  title = {Monotonically convergent optimization in quantum control using {Krotov's} method},
  journal = {Journal of Chemical Physics},
  volume = {136},
  pages = {104103},
  year = {2012}
}

@article{Kosloff1989,
  author = {Kosloff, Ronnie and Rice, Stuart A.},
  title = {Wavepacket dancing: Achieving chemical selectivity by shaping light pulses},
  journal = {Journal of Chemical Physics},
  volume = {90},
  number = {12},
  pages = {6947--6956},
  year = {1989}
}

@article{Shi1988,
  author = {Shi, Shuo and Woody, Austin and Rabitz, Herschel},
  title = {Optimal control of selective vibrational excitation in harmonic linear chain molecules},
  journal = {Journal of Chemical Physics},
  volume = {88},
  number = {2},
  pages = {6870--6883},
  year = {1988}
}

@article{Somloi1993,
  author = {Soml{\'o}i, J{\'a}nos and Kazakov, Vladimir A. and Tannor, David J.},
  title = {Controlled dissociation of {I}$_2$ via optical transitions between the {X} and {B} electronic states},
  journal = {Chemical Physics},
  volume = {172},
  number = {1},
  pages = {85--98},
  year = {1993}
}

@article{Nebendahl2009,
  author = {Nebendahl, Volker and H{\"a}ffner, Hartmut and Roos, Christian F.},
  title = {Optimal control of entangling operations for trapped-ion quantum computing},
  journal = {Physical Review A},
  volume = {79},
  pages = {012312},
  year = {2009}
}

@article{Muller2011,
  author = {M{\"u}ller, Markus M. and Reich, Daniel M. and Murphy, Michael and Yuan, Huan and Koch, Christiane P. and Whaley, K. Birgitta and Calarco, Tommaso},
  title = {Optimizing entangling quantum gates for physical systems},
  journal = {Physical Review A},
  volume = {84},
  pages = {042315},
  year = {2011}
}

@article{Rach2015,
  author = {Rach, Niels and M{\"u}ller, Markus M. and Calarco, Tommaso and Montangero, Simone},
  title = {Dressing the chopped-random-basis optimization: A bandwidth-limited access to the trap-free landscape},
  journal = {Physical Review A},
  volume = {92},
  pages = {062343},
  year = {2015}
}

@article{Goerz2014,
  author = {Goerz, Michael H. and Motzoi, Felix and Whaley, K. Birgitta and Koch, Christiane P.},
  title = {Charting the circuit {QED} design landscape using optimal control theory},
  journal = {npj Quantum Information},
  volume = {3},
  pages = {37},
  year = {2017}
}

@article{PalaoReichKoch2013,
  author = {Palao, Jos{\'e} P. and Reich, Daniel M. and Koch, Christiane P.},
  title = {Steering the optimization pathway in the control landscape using constraints},
  journal = {Physical Review A},
  volume = {88},
  pages = {053409},
  year = {2013},
  doi = {10.1103/PhysRevA.88.053409}
}

\end{document}